\documentclass[12pt]{iopart}
\usepackage{epsf}
\begin{document}

%

\newcommand{\be}{\begin{equation}}
\newcommand{\ee}{\end{equation}}
\newcommand{\bea}{\begin{eqnarray}}
\newcommand{\eea}{\end{eqnarray}}

%
%
%

\title{Vorticity and magnetic shielding in a type-II superconductor
}
\author{Marco Cardoso, Pedro Bicudo and Pedro D. Sacramento}
\address{Departamento de F\'{\i}sica and CFIF, Instituto Superior T\'ecnico,
Av. Rovisco Pais, 1049-001 Lisboa, Portugal}
\vspace{0.3cm}
\begin{abstract}
We study in detail, solving the Bogoliubov-de Gennes equations, the magnetic
field, supercurrent and order parameter profiles originated
by a solenoid or magnetic whisker inserted in a type-II superconductor. 
We consider solutions of different vorticities, $n$, in the various cases.  
The results confirm the connection between the vorticity, the internal currents
and the boundstates in a self-consistent way. 
The number of boundstates is given by the vorticity of the phase of the
gap function as in the case with no external solenoid. 
In the limiting case of an infinitely thin solenoid, like a
Dirac string, the solution is qualitatively different. The quasiparticle
spectrum and wave functions are a function of $n-n_{ext}$, where $n_{ext}$
is the vorticity of the solenoid. The flux is in all cases determined by
the vorticity of the gap function. 
\end{abstract}

\section{Introduction}

The effect of a magnetic field on a superconductor has attracted interest for a long time.
At small fields the superconductor is rigid to the external field but for a sufficiently
strong field, or a high enough electric current flowing through the material, superconductivity
is destroyed. In type-II superconductors there is a low critical field,
$H_{c1}$, above which the field lines penetrate the superconductor in the form of quantized
vortex lines. As the field increases, 
the vortex line density also increases until
the vortex cores overlap and the system becomes non-superconducting at the high critical
field, $H_{c2}$.
The external field is shielded by the appearance of compensating currents
that perfectly cancel the external field beyond a penetration length, $\lambda$. Another
length which is important is the coherence length, $\zeta$, which measures the range of the
establishment of the superconducting order parameter into the superconducting region.
Since many important superconductors, like the high-$T_c$ materials, are strong type-II
superconductors, the presence of external magnetic fields
implies proliferation of vortices and, therefore, their
influence has been thoroughly studied. 

In this work we consider the effects of a vortex induced by 
the insertion of a long solenoid or magnetic
whisker of very small width in the superconductor. 
Due to recent technological advances it is possible to construct magnetic systems
of nano size of the order of the coherence length or penetration length. We consider
that the inserted foreign system is long enough so that the field lines penetrate
the material (eventually through the effect of the vector potential) but that the field lines
return far from the superconducting film, such that no antivortices are created inside
the material. Solving the Bogoliubov-de Gennes equations we study in detail the
energy spectra and its consequences on the physical properties due to the
presence of the external solenoid, paying special attention to the magnetic
shielding due to the Meissner effect.

The problem of the quasi-particle states due to the presence of a vortex in a $s$-wave
superconductor was solved long ago both analytically \cite{caroli} and numerically \cite{Gygi}.
There are bound states localized in the vicinity of the vortex location and a continuum
of delocalized states. The case of a $d$-wave superconductor lead to some controversy
but it was established that the states are delocalized, consistently with a gapless
spectrum \cite{franz1}. Classifying the states in terms of the angular momentum
around the vortex line, allowed to determine that there is a branch of boundstates,
one for each angular momentum value \cite{Gygi}. The results were obtained looking for
an order parameter of the form $\Delta=\Delta_0 e^{-in\varphi}$ where
$\varphi$ is the polar angle and $n$ fixes the vorticity, chosen originally as $n=1$.
The core states are coherent superpositions of particle
and hole states and interpreted as being the result of constructive interference 
of multiple Andreev scattering
from the spatial variation of the order parameter \cite{rainer}.
Also, it was shown that the main contribution to the supercurrent is originated in these
states.
At very low temperatures the quantum nature of the bound states gains importance and
oscillations in the various quantities are quite pronounced \cite{hayashi}.
The quantum limit is obtained when the thermal width is smaller than the level
spacing and is reached when $T/T_c\leq 1/(k_F \zeta_0)$, where
$T_c$ is the critical temperature, $k_F$ is the Fermi momentum and $\zeta_0=v_F/\Delta_0$
is the coherence length. In the typical type-II layered superconductor $NbSe_2$ the
critical temperature is $T_c=7.2K$ and $k_F \zeta_0 \sim 70$. The quantum limit is
reached for rather low temperatures of the order of $50mK$, but for the high temperature
superconductors, where the critical temperature is high and the coherence length is small,
the quantum limit is reached for temperatures of the order of $10K$.
In this limit, Friedel-like oscillations are found in contrast to a Ginzburg-Landau
description. In particular the Kramer-Pesch effect \cite{kramer} 
where the vortex core size increases
with temperature, is not explainable in terms of a description in terms of normal
electrons. The oscillations in the various quantities are observed at sufficiently low
temperatures, irrespective of the size of $k_F \zeta_0$.

Even though as the field increases it is energetically favorable that many single
vortices appear, as compared to fewer vortices with higher units of flux \cite{dg}, 
the possibility of multi-flux vortices (or giant vortices) 
has been considered in particular systems and situations. 
In this case it has been predicted that there are $n$
branches of boundstates (where $n$ is the vorticity of the flux line) \cite{volovik}
and confirmed by several authors \cite{virtanen1}.
If formed, the doubly quantized vortex is metastable against dissociation into
singly-quantized vortices. In this case a counter-circulating current appears in the
core \cite{rainer} which is due to a bound state that appears below the Fermi level.
At high temperatures this state is depopulated, the current reversal disappears
and leads to a structure similar to the Ginzburg-Landau description.
The appearance of $n$ discrete branches of states for a vortex with vorticity $n$ should
be seen in Scanning Tunneling Microscopy (STM) measurements in the form of $n$ rows
of peaks as a function of the distance from the vortex core \cite{virtanen1}. 
Oscillations in
the current due to the multiple vortices were observed at low temperatures in a form that
is qualitatively different from results obtained from a Ginzburg-Landau
description \cite{virtanen2}. These oscillations are also observed in the order parameter.

Multiply quantized vortices have been observed in type-I superconductor films at high
magnetic fields \cite{hasegawa}. In high temperature superconductors having columnar
or large pointlike defects acting as pinning centers, multiply quantized vortices
may appear as well.
Giant-vortex states have also been observed in superconductors with
strong geometrical constraints, like in triangles or squares of sizes of the order of
the $\mu m$, using a scanning SQUID microscope \cite{nishio}. The possibility of
different vorticities leads to interesting features like the Little Parks effect where
oscillations of the critical temperature as the external field changes occur due
to transitions between states with different vorticities \cite{little}. 

In general, we may have a mixture of vortices with different winding numbers. Recently,
in a $NbSe_2$ traditional superconductor, where normal metal
islands of gold are inserted, a coexistence of strongly interacting multiquanta vortices
distributed in a lattice with interstitial single vortices, has
been observed \cite{karapetrov}. 
The case of a single normal dot inserted in a superconductor was also studied \cite{tanaka}.
The dot is quite small of a subnanometer scale. It was found that the energy
spectrum strongly depends on the number of flux quanta penetrating the dot and it
was shown that the number of branches corresponds to the number of flux quanta.
To guarantee that any vortices in the system would be in the dot, fields smaller than
$H_{c1}$ were used. It was shown that by increasing the dot size the vorticity
of the dot could be increased and therefore different states could be studied directly.

The vortex lines in general appear due to the application of an external
magnetic field typically homogeneous. However, we can as well consider the
presence of magnetic field lines that are due to a solenoid or a magnetic
rod inserted in the superconductor. Actually, the magnetic field lines do not
need to penetrate the superconductor itself, since what really matters is the
vector potential. It is the vector potential that appears in the Hamiltonian
of the system in the presence of a magnetic field \cite{dg}, as is well known.
This has been emphasized \cite{lyuksyutov1} 
considering a superconductor in the form of a cylindrical
shell of internal radius $R$ and width $a$ in the center of which is inserted
either a solenoid or a magnetic rod of radius $r$ smaller than $R$. In these
systems, considering the length of the cylinders very large compared to the
radius, the field lines will close far from the supercondutor and therefore
no field lines penetrate the cylinder. However, the vector potential due
to the flux contained in the transverse section is non-zero and the field
has the same effect on the supercurrents. 

We may as well consider other situations
in which there is an interplay between magnetic systems and superconductors.
Recently the situation of ferromagnet-superconductor hybrid systems has been
reviewed \cite{lyuksyutov2}. On one hand it has been proposed that we can
manipulate spin and charge in magnetic semi-conductors using superconducting
vortices, with applications in spintronics \cite{berciu}.
On the other hand we may consider the effect of the magnetic system on
the superconductor.
The magnetic film is separated from the superconducting layer to avoid suppression
of superconductivity by the proximity effect. This is attained introducing between
the two films a thin insulating layer.

A possibility to insert a magnetic field in the superconductor is through the
field originated by a magnetic dot. Depending on the strength of the magnetization
of the dot we may have single or multiple vortices. 
Using a London
theory it has been found that giant vortices occur when the dot size is small
enough \cite{erdin1} with a size of the order of $4.5 \zeta$. Otherwise the
energetically favorable situation is the presence of single quantized vortices \cite{erdin1}.
Considering an array of magnetic dots these originate vortices
in the superconductor which are preferably bound to the magnetic dots in a way that
is more advantageous than due to the usual defect pinning centers \cite{lyuksyutov3}.
In zero field the dots, which are far apart, are not coupled and are oriented
randomly. Therefore any other vortices that may not be bound to the dots or that
appear as fluctuations will feel the presence of a random magnetization
and will be in a resistive state. However, if a field is applied the dots align
and if field cooled the film may be superconducting. The ordered commensurate
state is then favored. The pinning turns out to be more effective than with
non-magnetic dots. 
Also, the critical current has the unusual property that it is increased 
with field.
Therefore introducing magnetic nanoparticles or nanorods one obtains
what are called frozen flux superconductors \cite{lyuksyutov4,lyuksyutov5}. 
In these systems a thermodynamically stable state of frozen flux lines is obtained
showing that the magnetic dots are more effective than other pinning centers.
This is important to control the maximum currents that may flow in the
superconductor. Additional flux lines created by an applied magnetic field need
to overcome big energy barriers in order to move. 
As mentionedabove larger systems were also considered such as two films, one magnetic and the
other superconductor, close by \cite{erdin2}. In this close vicinity the magnetic
field created by the supercurrents interacts with the magnetic subsystem. 
The interplay between the two systems leads to interesting magnetic structures
like the cryptoferromagnetic state \cite{bergeret1} and commensurability effects lead
to increase of critical current \cite{aladyshkin}. 

Since the total flux created
by the magnetic dot is zero if the superconducting film is large enough, one may
expect the presence of vortices and anti-vortices, due to the dipole field
of the dot. 
In a type-II
superconductor the interaction between two vortices is repulsive and the interaction
between a vortex and an anti-vortex is attractive. In studies close to the critical
line it has been found that the antivortices may coexist with the vortices but away
from the vicinity of the critical line they disappear. However, in type-I superconductors
the interaction between a vortex and an anti-vortex is repulsive (and the interaction
between two vortices is attractive). Therefore it is possible that structures
with vortices and antivortices may appear in type-I systems. Also, small systems
show confinement effects on the vortices: due
to the strong increase of the kinetic energy term near the frontier, it is favorable for the
system to be in the superconducting state close to the borders. Therefore it is not
favorable for a vortex to approach the borders of the system. As a consequence they tend to
be close together near the center, particularly if the system size is very small.
Considering a type-I system with a triangular symmetry it has been shown that indeed 
stable vortex-antivortex structures are possible \cite{misko1}. A rich 
variety of vortex structures has been studied recently \cite{milosevic}.

All of these results show the importance of the study of complex structures of
vortices and their interplay with magnetic structures. Recently it was
stressed that the important characteristic that determines the boundstates is the
winding of the phase \cite{berthod}. The detailed form of the order parameter in the
vicinity of the vortex core is not so relevant. Performing a non-self-consistent
study of the spectrum it was found that the suppression of the gap function
has a minor role. The important feature is the winding of the phase.
The supercurrent acts in non-symmetrical way on the particle and hole parts
of the quasiparticles. It tends to decrease the angular momentum of the
particle part and to increase the angular momentum of the hole part. 

In this work we consider
the same problem from a different point of view. We perform a self-consistent
solution of the influence of a very long solenoid on the properties
of the system. We consider different possible vorticities and study
how the system responds to the external perturbation.
The internal field must adjust
itself to the solution chosen according to the external field exerted by the solenoid.
The total magnetic flux is fully determined by the choice
of the angular momentum of the gap function and the value of $n$ determines
the vorticity of the vortex solution. This may be a single or a multiple vortex.
Depending on the relation of the value of $n$ and the value of the external
field, the internal currents will create fields that compensate, undercompensate
or overcompensate the external field. The various situations lead to different
spectral structures depending on the width of the solenoid. 
These in turn originate different structures for the
internal field and supercurrents generated. 
The limit of a very thin solenoid is qualitatively different.
It is shown that when the vorticity
chosen equals the unit of external flux the currents generated vanish and no
bound states appear. Otherwise the currents may be positive, and branches of
boundstates with positive energies appear or the currents are negative and branches
of negative energies appear. These results confirm the link between the boundstates
and the internal currents in a self-consistent way.
In the case of a finite width solenoid the number of boundstates equals
the vorticity of the gap function and is insensitive to the external field.

\section{Method}

\subsection{Bogoliubov-de Gennes equations}
Consider the Bogoliubov-de Gennes equations (BdG)
\bea
\left[ \frac{1}{2m}(\mathbf{p}-e \mathbf{A})^2 + U(\mathbf{r}) - E_F \right] u_i(\mathbf{r}) 
+ \Delta(\mathbf{r}) v_i(\mathbf{r})  &=& E_i u_i(\mathbf{r}) \nonumber \\
-\left[ \frac{1}{2m}(\mathbf{p}+e \mathbf{A})^2 + U(\mathbf{r}) - E_F \right] v_i(\mathbf{r}) 
+ \Delta^{*}(\mathbf{r}) u_i(\mathbf{r})  &=& E_i v_i(\mathbf{r})
\eea
where $U(\mathbf{r})$ is an external potential,
$\mathbf{A}(\mathbf{r})$ is the vector potential and where we consider
$s$-wave pairing, for simplicity. $\Delta(\mathbf{r})$ 
is the pairing function given by
\begin{equation}
\Delta(\mathbf{r}) = g \sum_{0 < E_i \leq \hbar \omega_D} u_i(\mathbf{r}) v_i^{*}(\mathbf{r})[1 - 2f(E_i)]
\end{equation}
Here $f(E_i)$ is the Fermi-Dirac distribution. 
The vector potential is given by Maxwell's equations
\begin{equation}
\nabla \times \mathbf{B} = \nabla \times \nabla \times \mathbf{A}
 = \frac{4 \pi}{c}\mathbf{J}_{total}
\end{equation}
which, in the Coulomb gauge ( $\nabla . \mathbf{A} = 0$ ), is given by 
\begin{equation}
\nabla^2 \mathbf{A} = - \frac{4\pi}{c} \mathbf{J}_{total}
\end{equation}
The current density originated in the supercurrents is obtained self-consistently by
\bea
\mathbf{J}(\mathbf{r}) &=& \frac{e\hbar}{2im} \sum_i f(E_i) u_i^{*}(\mathbf{r}) 
\left[ \nabla - \frac{ie}{\hbar c}\mathbf{A}(\mathbf{r}) \right] u_i(\mathbf{r}) \nonumber \\
&+& [(1-f(E_i)] v_i(\mathbf{r}) \left[ \nabla - 
\frac{ie}{\hbar c}\mathbf{A}(\mathbf{r})\right] v_i^{*}(\mathbf{r}) - c.c.
\eea

\subsection{Diagonalization of the Hamiltonian}
We assume no dependence along the
axis of the vortex line ($z$-axis) and cylindrical symmetry
both of the superconductor and of the potential $U(\bf r)$. 
Let us take the order parameter in the form
\begin{equation}
\Delta(\mathbf{r}) = \Delta(\rho) e^{-i n \varphi}
\end{equation}
This form describes a magnetic flux equal to $n$ flux quanta
( $\Phi= n \Phi_0 = n \frac{hc}{2e}$ ).

The wave functions $u_i$ and $v_i$ are expanded in a way similar to ref. \cite{Gygi} 
\begin{equation}
u_i(\mathbf{r}) = \sum_{\mu,j} c_{\mu,j}^i \phi_{j,\mu-1/2} e^{i(\mu-1/2)\varphi}
\end{equation}
\begin{equation}
v_i(\mathbf{r}) = \sum_{\mu,j} d_{\mu,j}^i \phi_{j,\mu-1/2+n} e^{i(\mu-1/2+n)\varphi}
\end{equation}
where the basis functions are
\begin{equation}
\phi_{jm}(\rho) = \frac{\sqrt{2}}{RJ_{m+1}(\alpha_{jm})}J_{m}
\left(\alpha_{jm}\frac{\rho}{R}\right) 
\end{equation}
The system is placed in a cylinder of radius $R$.
Given the azymuthal symmetry of the system, neither $\Delta(\rho)$
nor $\mathbf{A}$ depend on $\varphi$. Therefore the Hamiltonian
may be diagonalized separately for each value of the angular
momentum $\mu$. The functions $J_m$ are the spherical Bessel functions and
$\alpha_{jm}$ is the $j^{th}$ zero of the Bessel function of order $m$.
The set of values of the angular momentum is given by $\mu=\pm (2l+1)/2$ where
$l=0,1,2,\cdots$.
The terms $(\mu-1/2)$ and $(\mu-1/2+n)$ may be rewritten in a more
symmetrical way like $(\mu^{\prime}-n/2)$ and $(\mu^{\prime}+n/2)$ if
$\mu^{\prime}$ is half-odd integer if $n$ is odd and $\mu^{\prime}$ is
integer if $n$ is even.

For each eigenvalue $E_i$, we have a single value of $\mu$ and it is enough
to diagonalize the matrix, defined in the subspace of the zeros of the Bessel
function,
\begin{equation}
\left( \begin{array}{cccc} 
		  T^{-} & \Delta \\
		  \Delta^T & T^{+} \\
		  \end{array}
		  \right)
\left( \begin{array}{c} 
		  c_{\mu}^i  \\
		  d_{\mu}^i \\
		  \end{array}
		  \right)		
= E_i
\left( \begin{array}{c} 
		  c_{\mu}^i  \\
		  d_{\mu}^i \\
		  \end{array}
		  \right)				   
\end{equation}
where
\bea
T_{j j'}^{-} &=& - \frac{\hbar^2}{2m} \frac{\alpha_{j,\mu-1/2}^2}{R^2} 
\delta_{jj'} - (\mu - 1/2) \frac{e}{\hbar c} I_1^- \nonumber \\ 
&-& \frac{e^2}{\hbar^2 c^2}I_2^- + E_F
\eea
\bea
T_{j j'}^{+} &=& + \frac{\hbar^2}{2m} \frac{\alpha_{j,\mu-1/2+n}^2}{R^2} 
\delta_{jj'} - (\mu-1/2+n) \frac{e}{\hbar c} I_1^+ \nonumber \\ 
&+& \frac{e^2}{\hbar^2 c^2}I_2^+ - E_F
\eea
with
\begin{equation}
I_1^- = \int_{0}^{R} \phi_{j,\mu - 1/2}(\rho) \frac{A_\varphi(\rho)}
{\rho} \phi_{j',\mu - 1/2}(\rho) \rho d\rho 
\end{equation}
\begin{equation}
I_1^+ = \int_{0}^{R} \phi_{j,\mu-1/2+n}(\rho) \frac{A_\varphi(\rho)}
{\rho} \phi_{j',\mu-1/2+n}(\rho) \rho d\rho 
\end{equation}
and
\begin{equation}
I_2^- = \int_{0}^{R} \phi_{j,\mu-1/2}(\rho) A_\varphi(\rho)^2 
\phi_{j',\mu- 1/2}(\rho) \rho d\rho 
\end{equation}
\begin{equation}
I_2^+ = \int_{0}^{R} \phi_{j,\mu -1/2+n}(\rho) A_\varphi(\rho)^2 
\phi_{j',\mu - 1/2+n}(\rho) \rho d\rho
\end{equation}
Also we have
\begin{equation}
\Delta_{jj'} = \int_{0}^{R} \phi_{j,\mu - 1/2}(\rho) 
|\Delta(\rho)| \phi_{j',\mu - 1/2+n}(\rho) \rho d\rho
\end{equation}
It is important to note that the symmetry of the BdG equations
\begin{equation}
u_i(\mathbf{r}) \rightarrow v_i^{*}(\mathbf{r})
\end{equation}
\begin{equation}
v_i(\mathbf{r}) \rightarrow -u_i^{*}(\mathbf{r})
\end{equation}
\begin{equation}
E_i \rightarrow - E_i
\end{equation}
allows to reduce the solution to the positive values of $\mu$.
We obtain the eigenvectors and eigenvalues for negative values of $\mu$ using the
above symmetry. 

\subsection{Calculation of the vector potential}

Consider a general situation where the total vector potential is given
by the sum of an external potential and the internal vector potential
originated on the supercurrents. Then we can write that
\be
\vec{A}=\vec{A}_{ext} + \vec{a}
\ee
where $\vec{a}$ is the internal gauge potential. Then naturally we can write
that
\bea
\Phi &=& \Phi_{a} + \Phi_{ext} \nonumber \\
\vec{B} &=& \vec{\nabla} \times \vec{A} = \vec{\nabla} \times \vec{a} + \vec{B}_{ext} \nonumber
\\
\nabla^2 \vec{A} &=& \nabla^2 \vec{a} + \nabla^2 \vec{A}_{ext} = \nabla^2 \vec{a}
\eea
since $\nabla^2 \vec{A}_{ext}=0$, except at the region where the external currents are non-zero.
Therefore we have to solve the equation
\be
\nabla^2 \vec{a} = -\frac{4\pi}{c} \vec{J}
\ee
Due to the quantization condition the total flux is given by
\be
\Phi = \Phi_a+ \Phi_{ext} = n \Phi_0
\ee
Therefore we will be considering situations where the supercurrents will generate
internal magnetic fluxes such that
\be
\Phi_a = (n-n_{ext}) \Phi_0
\ee
where $\Phi_{ext}=n_{ext} \Phi_0$ (we take $n_{ext}$ as a real parameter).

Let us then start from the equation
\begin{equation}
\frac{1}{\rho}\frac{\partial}{\partial \rho}\left( \rho 
\frac{\partial a_\varphi}{ \partial \rho}\right)    
-  \frac{a_{\varphi}}{\rho^2}
= -\frac{4\pi}{c} J_{\varphi}
\end{equation}
Defining $a_{\varphi} = F(\rho) / \rho$, 
Poisson's equation reduces to
\begin{equation}
\frac{\partial^2 F}{\partial \rho^2} - \frac{1}{\rho} \frac{\partial F}{\partial \rho} = 
-\frac{4\pi}{c}J_\varphi \rho
\end{equation}
Since the current is given by
\bea
& &J_\varphi(\rho,z) = \frac{1}{m} \sum_{i, \mu > 0} f(E_i) |u_i(\mathbf{r})|^2 \left[ \frac{\mu-1/2}{\rho} 
- \frac{e}{\hbar c}A_\varphi(\rho,z) \right] \nonumber \\
&+&  [ 1 - f(E_i) ] |v_i(\mathbf{r})|^2 
\left[ -\frac{\mu-1/2+n}{\rho} - \frac{e}{\hbar c}A_\varphi(\rho,z) \right] \nonumber \\
\eea
we can make the decomposition
\begin{equation}
-\frac{4\pi}{c}J_\varphi \rho = K(\rho) + \beta(\rho)F(\rho)
\end{equation}
with
\bea
& & K(\rho) = 
-\frac{4\pi}{mc} \sum_{i} f(E_i) |u_i|^2 (\mu-1/2) \nonumber \\
& & - [1-f(E_i)] |v_i|^2 (\mu-1/2+n) 
\nonumber \\
& & + \frac{4\pi}{c^2m} \rho A_{\varphi}^{ext}(\rho) 
\sum_i  \left\{ f(E_i) |u_i|^2 + [1-f(E_i)] |v_i|^2 \right\} \nonumber \\
\eea
and
\begin{equation}
\beta(\rho) = -\frac{4\pi}{mc} \frac{|e|}{\hbar c}\sum_{i} f(E_i) |u_i|^2 + 
[ 1 - f(E_i)]  |v_i|^2                              
\end{equation}
Therefore we get
\begin{equation}
\frac{\partial^2 F}{\partial \rho^2} - \frac{1}{\rho} \frac{\partial F}{\partial \rho} 
=
K(\rho) + \beta(\rho) F(\rho)
\end{equation}
Discretizing this equation
$$
\frac{\partial^2F}{\partial \rho^2} \rightarrow \frac{ F_{i+1} - 2F_{i} + F_{i-1} }{ a^2 }
$$
$$
\frac{\partial F}{ \partial \rho } \rightarrow \frac{ F_{i+1} - F_{i-1} }{ 2 a }
$$
where $a = R /( Nr + 1)$, and $N_r+1$ is the number of points.
We get
\begin{equation}
\frac{ F_{i+1} - 2F_{i} + F_{i-1} }{ a^2 }
- \frac{ F_{i+1} - F_{i-1} }{ 2 \rho a }
= K_i + \beta_i F_i
\end{equation}
The boundary conditions are such that $F(0) = 0$
( $a_\varphi$ does not diverge at the origin)
and $\frac{\partial F}{ \partial \rho}\big|_{\rho=R} = 0$ ( $\mathbf{B} = 0$ outside). 
In the first case it is enough to take $F_0 = 0$. In the second case we have
$F_{N+1} - F_{N} = 0$. In the outside boundary, $i = N_r$, we have
\begin{equation}
\frac{-F_N + F_{N-1}}{a^2} - \frac{F_N - F_{N-1}}{2a(R-a)} = K_N + \beta_N F_N
\end{equation}
The system is therefore reduced to a tridiagonal system of equations.

The equations are solved self-consistently choosing appropriately the first
iteration. The solution converges after a few iterations usually less than
$10$ iterations.

\section{Results}

\begin{figure}
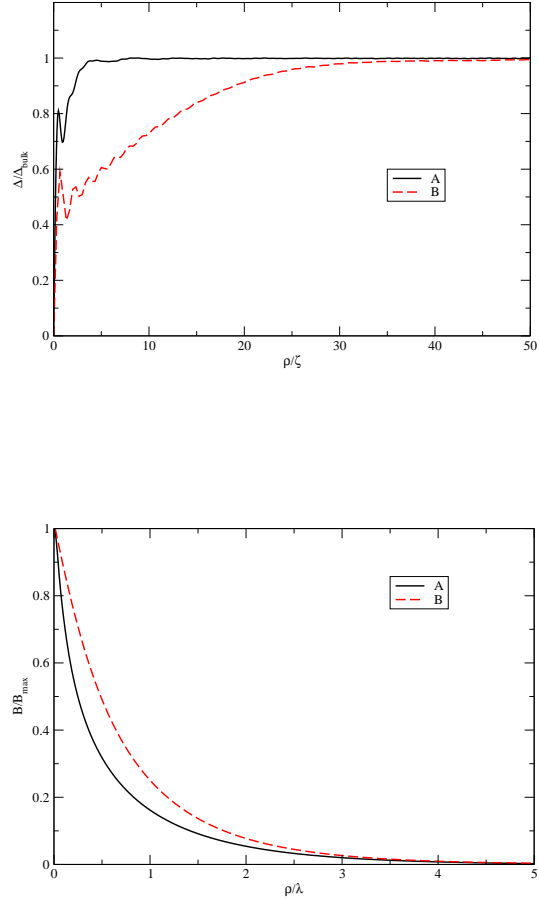

\epsfxsize=7.0cm \epsfysize=7.0cm
\centerline{\epsffile{Fig1a.eps}} \epsfxsize=7.0cm
\epsfysize=7.0cm \centerline{\epsffile{Fig1b.eps}}
\begin{center}
\caption{Comparison of the gap function and the magnetic field for the two
sets of parameters A and B, defined in the text, at a very low temperature. 
We take $n=1$ and no external solenoid ($n_{ext}=0$). The vertical axis
is scaled to the bulk value of the gap function and the value of the field
at the origin, respectively,  and the horizontal axis
is scaled to the coherence length and the penetration length, respectively, of each
set of parameters.
}
\end{center}
\label{fig1}
\end{figure}

The solution of the BdG equations gives full information about the superconductor
within BCS theory. 
The equations are solved in atomic units where
$m=\hbar=e=1$, $c=\frac{1}{\alpha}\approx 137$.
We consider two sets of parameters. One set (designated set A) 
corresponds to an extreme case in the quantum limit 
and another set (designated set B) corresponds approximately to the parameters suitable for the
traditional superconductor $NbSe_2$. In the first case we consider parameters such that
$ g = 0.8 $, $ E_F = 0.5 $, $ \omega_D = 0.25 $, 
$ R = 80-250 $, $ n_r = 1001 $, $ n_j = 200 $ and $ n_{\mu} = 600 $.
According to these parameters we have that $\zeta_0=v_F/\Delta_0 \sim 7.69$,
$k_F \zeta_0 \sim 7.7$. The critical temperature is of the order of $0.1$ (all of these
numbers are in atomic units). The second set of parameters is given by
$ g = 0.31 $, $ E_F = 37.2 meV $, $ \omega_D = 3 meV $, 
$ R = 10000 $, $ n_r = 1001 $, $ n_j = 200 $ and $ n_{\mu} = 600 $.
According to these parameters we have that $\zeta_0=v_F/\Delta_0 \sim 876$,
$k_F \zeta_0 \sim 65$. The critical temperature is of the order of $8K$.
(Recall that one atomic unit of distance is $\sim 0.5$ Angstrom, and that
one atomic unit of energy is $\sim 27 eV$).
In spite of the large difference between the two sets of parameters we will
see that many of the results are smilar. In the first case we consider temperatures
that are rather small of the order of $T/T_c \sim 0.001$ which corresponds to
the quantum limit regime (in the case of $NbSe_2$ this regime sets at a
temperature of the order of $50mK$). The scale of the energies is such that
in the first case $\Delta_0 \sim 3.53 eV$ while in the second set it is of the
order of $\Delta_0 \sim 1.2 meV$.

\begin{figure}
\epsfxsize=8.0cm
\epsfysize=8.0cm
\centerline{\epsffile{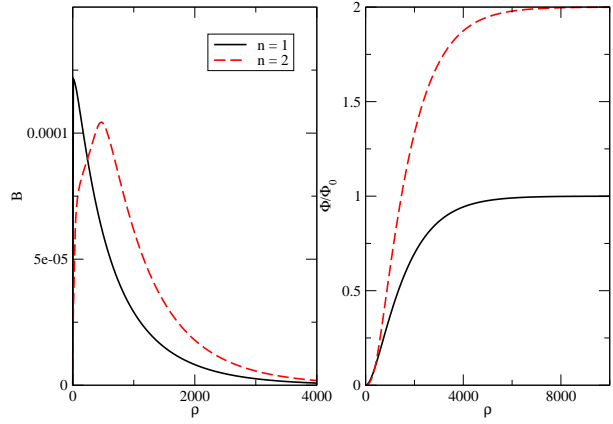}}
\caption{Magnetic flux and magnetic field for $n = 1$ and $n = 2$ in $NbSe_2$ (set B), 
with $T = 1K$ and no external solenoid.
}
\label{fig2}
\end{figure}

\subsection{$\mathbf{A^{ext}}(\mathbf{r})=0$}

We begin by reviewing the case where there is no external potential 
$U(\mathbf{r})$ and no explicit
$\mathbf{A^{ext}}(\mathbf{r})$, in addition to a vortex line characterized by the
vorticity $n$.
In the first iteration we choose
$\Delta(\mathbf{r})=\Delta_0 \tanh{ \frac{\rho}{\zeta} }$
and $\mathbf{A}(\mathbf{r})=\mathbf{A^{ext}}(\mathbf{r})=0$.
This is the usual way to study the presence of a vortex line induced in the
superconductor by the action of an external field, probably uniform.
The energy spectra for different vorticities show
the characteristic energy gap considering $s$-wave symmetry. 
For each angular momentum value there is a set of energy eigenvalues
(in number given by the basis set of Bessel function zeros) with positive and negative
values. The negative values give information about the positive energy values for the
negative angular momenta, as explained above.
The $n=0$ solution corresponds to the absence of a vortex and is therefore the
result expected in the bulk of the superconductor away from the vicinity of any vortex.
For $n \geq 1$ the solution corresponds to a vortex with increasing vorticity.
In these cases a set of bound states appears in addition to the continuum. The
number of states for each angular momentum value is given by the vorticity.
For $n=1$ the boundstates are all positive (the Caroli, de Gennes and Matricon
states \cite{caroli}) but for $n=2$, and $n=3$ some of the boundstates have negative
energies.

The range over which the gap function
reaches its bulk value defines the coherence length. For $n=1$ the gap function is
always positive but for $n=2$ and $n=3$ it has a node. 
This is associated with the appearance of negative energy boundstates.
The oscillations in the
various curves are eliminated if the temperature is increased and in general the coherence
length increases with the temperature (Kramer-Pesch effect). Also the coherence length
increases with the vorticity.

The magnetic field is shielded in the superconductor. 
In the case of a $n=1$ vortex the magnetic field decreases
monotonically evolving in the bulk to an exponential decay, which defines the penetration
length, $\lambda$. As the vorticity increases the magnetic field profile is no longer
monotonic but there is a maximum at a finite distance from the vortex. Also
the penetration length increases with the vorticity, as does the coherence length.
In Fig. 1 we compare the normalized gap functions and the normalized 
magnetic field for the two sets of parameters considered. 
In the first case we estimate a coherence length of the order of $\zeta \sim 2$ and
in the second case $\zeta \sim 100$. From the decay of the magnetic field we estimate
that in the first case $\lambda \sim 43$ while in the second case $\lambda \sim 790$.
This implies ratios of the order of $\lambda / \zeta \sim 23$ and 
$\lambda / \zeta \sim 8$, respectively.

In order for the magnetic field to be shielded an internal field is generated in the
superconductor to compensate the external field. This field is generated by supercurrents
that appear in the superconductor due to the motion of the Cooper pairs. 
At the origin there is
no current since the total field equals the external field and no shielding takes place.
Considering first the $n=1$ case the current increases to compensate the external field
and then decreases exponentially. The cases of higher vorticity are more complex \cite{rainer}.
The current is negative near the origin, then becomes positive and follows the same
trend as for the singly quantized vortex. 
The increased winding of the phase around the vortex induces in the immediate vicinity
of the vortex core currents that have opposite circulation.
In Fig. 2 we show, for completeness, the magnetic field and the flux for the cases of $n=1$ and
$n=2$ for $NbSe_2$ (parameters B).

\begin{figure}
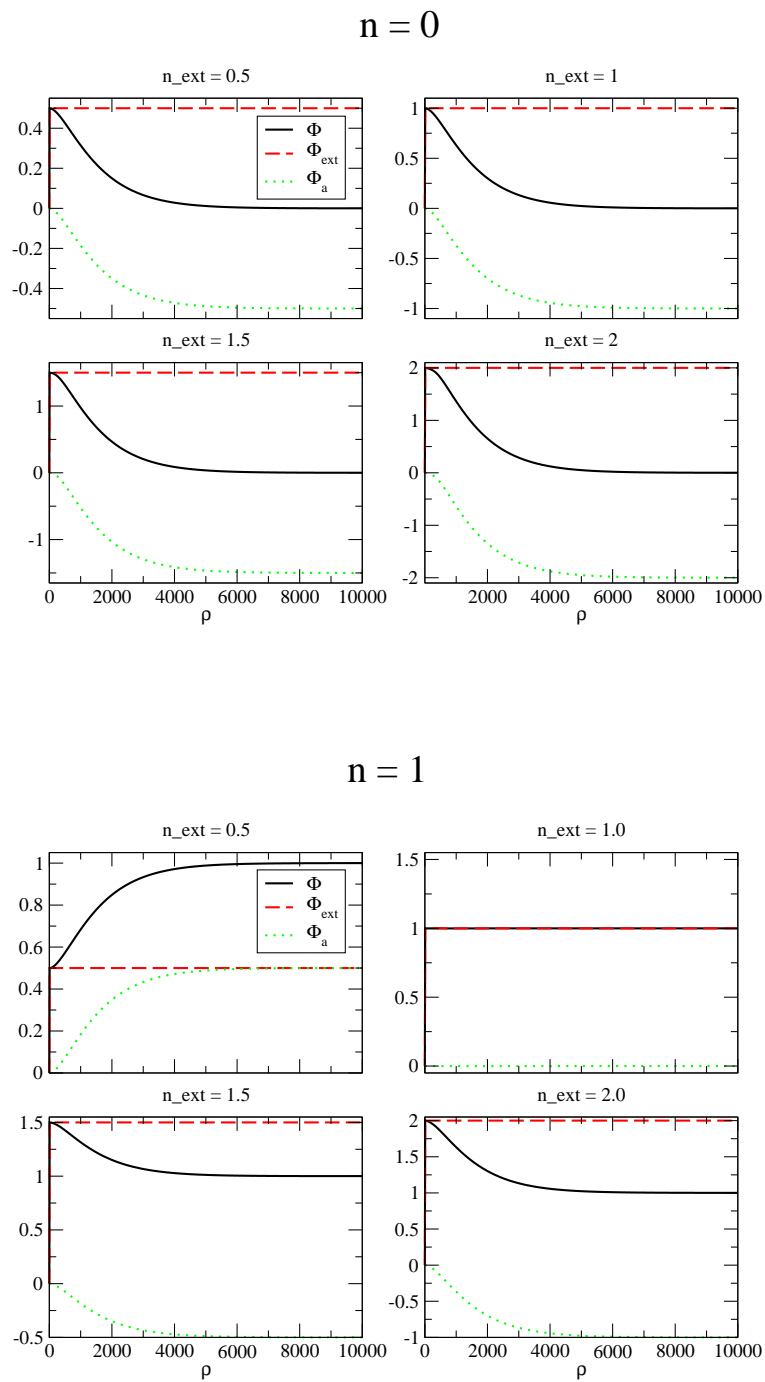

\epsfxsize=10.0cm \epsfysize=10.0cm
\centerline{\epsffile{Fig3a.eps}} \epsfxsize=10.0cm
\epsfysize=10.0cm \centerline{\epsffile{Fig3b.eps}}
\begin{center}
\caption{Magnetic flux (internal, external and total), for $n = 0$ (top panel)
and for $n = 1$ (bottom panel) and different $n_{ext}$ in $NbSe_2$ (set B), with $T = 1K$.
}
\end{center}
\label{fig3}
\end{figure}

\subsection{External field line: solenoid of negligible width}

Consider now that in addition or instead of the uniform external field we insert
in the superconductor a solenoid of negligible width. 
A current may be applied and the solenoid (assumed infinitely long)
creates a magnetic field line of negligible width inside it and a
zero field outside. 
The solution of the BdG equations depends both on the vorticity of the gap
function and on the total field through the vector potential $\vec{A}=\vec{a}+
\vec{A}_{ext}$. 
The external vector potential due to the thin solenoid is given in this limitting case by
\be
A_{ext}^{\varphi} = \frac{n_{ext} \Phi_0}{2 \pi \rho}
\ee
However, in the numerical solution we replace $\rho \rightarrow \tilde{\rho}$, 
where $\tilde{\rho}=\sqrt{\rho^2+\delta^2}$ is introduced to regularize the vector potential
at the origin ($\delta$ is infinitesimal).
In most cases the term which includes the vector
potential may be neglected in the BdG equations because it is either negligible or very small
and the main influence of the vector potential is in the expression for the current.
If there is no external vector potential the internal vector potential (and consequently
the total vector potential) vanishes at the axis (like $\sim \rho$).
However, if the external potential has a singular form this term is important
and must be taken into account. We note that we include the vector potential term in the
BdG equations even though its contribution is small \cite{Gygi}.

When the external solenoid is inserted and a field generated, one expects the
right solution to be the one corresponding to a vorticity matching the total
field. One may however look for other solutions of the BdG equations which
eventually will have higher free energies as compared to the right solution.
We will in the following consider different possible solutions of the BdG equations
even though only one is the lowest free energy.

Consider again the BdG equations. Assuming that 
$\mathbf{A}(\mathbf{r}) = A_{\varphi}(\rho) \hat{e}_{\varphi}$
with 
\begin{equation}
A_{\varphi} = n_{ext} \frac{\hbar c}{2 e \rho} + a_{\varphi}
\end{equation}
and making the general assumption of azimutal symmetry, we can write the eigenvectors as
\begin{equation}
u_i = g_{\mu' j}(\rho) e^{i (\mu' - n/2) \varphi}
\end{equation}
\begin{equation}
v_i = h_{\mu' j}(\rho) e^{i (\mu' + n/2) \varphi}
\end{equation}
Then we have
\bea
H_e u_i &=& 
 \Big(
 -\frac{\hbar^2}{2m} 
 \Big( 
 \frac{1}{\rho} \frac{d}{d \rho} \big( \rho \frac{d g_{\mu' j}}{d \rho} \big)
- \frac{(\mu' - n/2)^2}{\rho^2} g_{\mu' j}
- \frac{2 (\mu' - n/2) e}{\hbar c} \frac{a_\varphi}{\rho} g_{\mu' j} \nonumber \\
&-& \frac{ (\mu' - n/2) n_{ext} }{ \rho^2} g_{\mu' j} 
 - \frac{n_{ext}^2}{4 \rho^2} g_{\mu' j}
 - \frac{n_{ext} e}{\hbar c} \frac{a_{\varphi}}{\rho} g_{\mu' j}
 - \frac{e^2}{\hbar^2 c^2} a_{\varphi}^2 g_{\mu' j}
 \Big) \nonumber \\
 &-& E_F g_{\mu' j}
 \Big)
 e^{i ( \mu' - n/2) \varphi}
\eea
Now, grouping and simplifying the terms 
the expression simplifies to
\bea
 H_e u_i &=& 
  \Big(
 -\frac{\hbar^2}{2m} 
 \Big( 
 \frac{1}{\rho} \frac{d}{d \rho} \big( \rho \frac{d g_{\mu' j}}{d \rho} \big)
- \frac{(\mu' - n_a/2)^2}{\rho^2} g_{\mu' j}
- (2 \mu' - n_a) \frac{e}{\hbar c} \frac{a_\varphi}{\rho} g_{\mu' j} \nonumber \\
&-& \frac{e^2}{\hbar^2 c^2} a_{\varphi}^2 g_{\mu' j}
 \Big)  
- E_F g_{\mu' j}
  \Big)
 e^{i ( \mu' - n/2) \varphi}
\eea
where $n_a=n-n_{ext}$.
For $H_e^* v_i$, the expression can be obtained from this one, 
by replacing $u_i$ by $v_i$ and $\mathbf{A}$ by $-\mathbf{A}$. 
The first replacement is equivalent to making
\begin{equation}
g_{\mu' j} \rightarrow h_{\mu' j}, 
n \rightarrow - n
\end{equation}
and the second is equivalent to
\begin{equation}
n_{ext} \rightarrow - n_{ext}, 
a_{\varphi} \rightarrow - a_{\varphi}
\end{equation}
So, $H_e^* v$ is given by
\bea
 H_e^* v_i &=& 
  \Big(
 -\frac{\hbar^2}{2m} 
 \Big( 
 \frac{1}{\rho} \frac{d}{d \rho} \big( \rho \frac{d h_{\mu' j}}{d \rho} \big)
- \frac{(\mu' + n_a/2)^2}{\rho^2} h_{\mu' j}
+ (2 \mu' + n_a) \frac{e}{\hbar c} \frac{a_\varphi}{\rho} h_{\mu' j} \nonumber \\
&-& \frac{e^2}{\hbar^2 c^2} a_{\varphi}^2 h_{\mu' j}
 \Big) 
- E_F h_{\mu' j}
  \Big)
 e^{i ( \mu' + n/2) \varphi}
\eea
So the explicit dependence on $n$ and $n_{ext}$, comes only through $n_a$.

The same choices for the vector potential and for the eigenfunctions
leads to the current in the form 
\bea
J_\varphi &=& 
-\frac{e \hbar}{m} \sum_{\mu' j}
f(E_{\mu' j}) g_{\mu' j}^2 \Big( \frac{\mu' - n_a / 2}{ \rho } + 
\frac{e}{\hbar c} a_{\varphi} \Big)  \nonumber \\
&+& \big(1-f(E_{\mu' j}) \big) h_{\mu' j}^2 \Big( \frac{-\mu' - n_a / 2}{ \rho } + 
\frac{e}{\hbar c} a_{\varphi} \Big)
\eea
Also, the Maxwell equation becomes
\bea
\frac{1}{\rho} \frac{\partial}{\partial \rho} \big( \rho 
\frac{a_\varphi}{\partial \rho} \big) - \frac{a_{\varphi}}{\rho^2}
&=&
-\frac{4 \pi e \hbar}{m c} \sum_i
f(E_{\mu j}) g_{\mu j}^2 \Big( \frac{\mu - n_a / 2}{ \rho } 
+ \frac{e}{\hbar c} a_{\varphi} \Big) \nonumber \\ 
&+& \big(1-f(E_{\mu j}) \big) h_{\mu j}^2 \Big( \frac{-\mu - n_a / 2}{ \rho } 
+ \frac{e}{\hbar c} a_{\varphi} \Big)
\eea
So neither the BdG equations, nor the Maxwell equation depends on $n$ or $n_{ext}$, separetely, 
but rather on the difference $n_a = n - n_{ext}$. 
The solution of the spectrum, eigenfunctions and all quantities derived from these
are only a function of $n_a$. The singular term of the external field line has the
effect of renormalizing the states of the Bessel function basis if the external
flux is a multiple of the quantum of flux. In particular, when $n-n_{ext}=n_a=0$ the
system has an effective zero vorticity. However, the total flux is determined by $n$,
as we will see.

If we consider the combined transformations $n_a \rightarrow -n_a$, $a_{\varphi}
\rightarrow -a_{\varphi}$ and $\mu' \rightarrow -\mu'$ the equations are
invariant. This is equivalent, as we saw before, to make the combined transformations
$n_a \rightarrow -n_a$, $a_{\varphi} \rightarrow -a_{\varphi}$, 
and $E_i \rightarrow -E_i$, $g \rightarrow h$ and $h \rightarrow -g$. Under these
transformations the gap function remains invariant and the current changes sign, as
expected.

In Fig. 3 we plot the flux as a function of 
distance for two different vorticities
$n=0$ and $n=1$ for different applied fields. The total field is determined
by the choice of vorticity of the solution: in the case of $n=0$ the total flux in the
bulk vanishes and in the case of $n=1$ the total flux in the bulk equals a quantum of flux.
The external field is determined by the current that goes through the infinitely thin
solenoid. This is parametrized by the flux $\Phi_{ext}=0.5,1,1.5,2 \Phi_0$.
The fully self-consistent solution of the BdG equations yields the internal vector
potential $\vec{a}$ generated by the supercurrents carried by the quasiparticles.
The flux originated by the internal field compensates the external flux in order
to give the correct total flux fixed by the chosen vorticity, $n$. As we can see
in Fig. 3, for all values of the external field, the internal field is negative in order
to compensate and give a zero total flux in the bulk. In the case of $n=1$ the value 
of $\Phi_{ext}=0.5 \Phi_0$ is smaller than the total flux and the internal vector
potential is positive. For $\Phi_{ext}=\Phi_0$ the internal field is zero and for
the higher values of $\Phi_{ext}$ the internal field is again negative, as explained above.

\begin{figure}
\epsfxsize=6.0cm \epsfysize=6.0cm \centerline{\epsffile{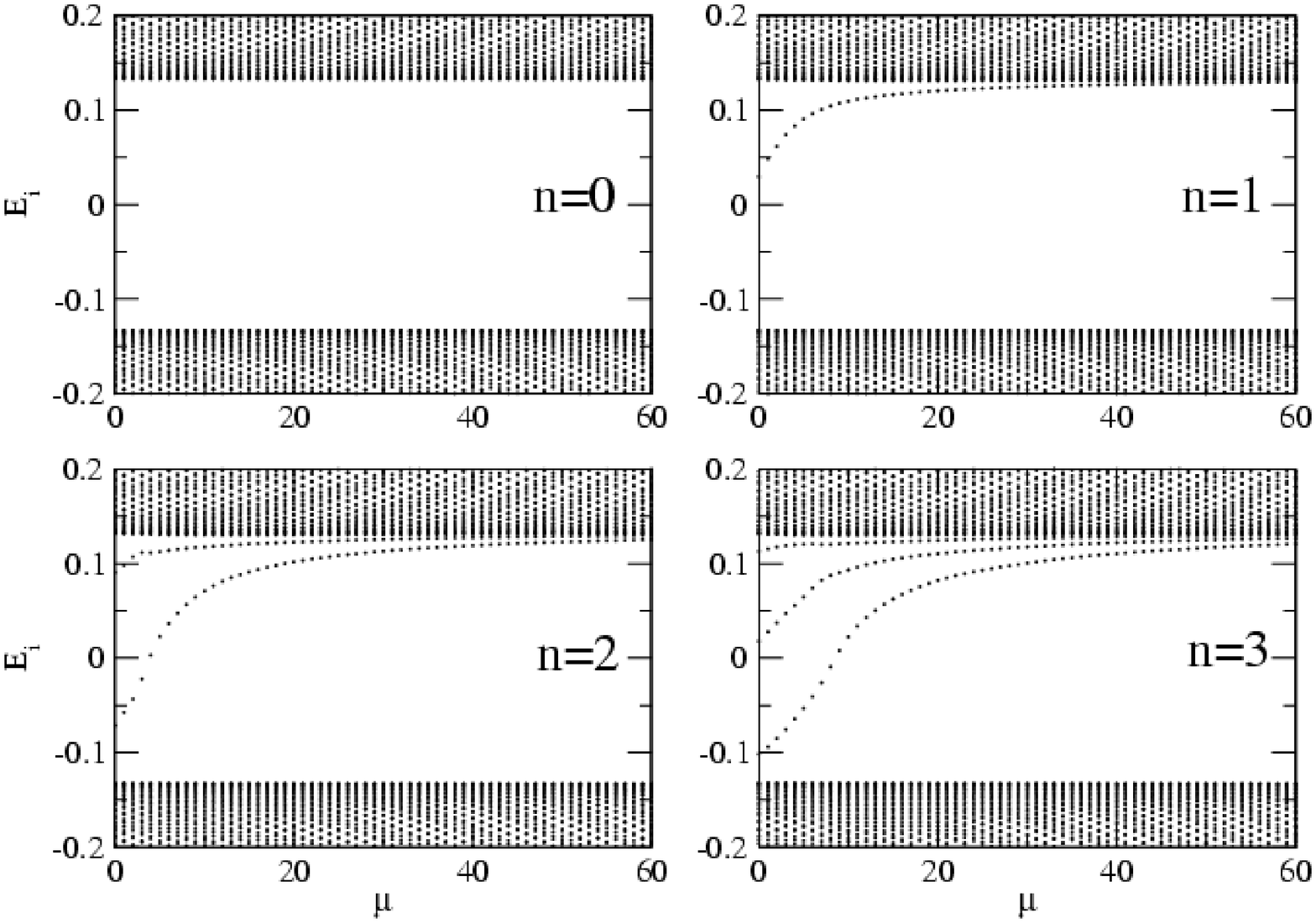}} 
\vspace{1cm} 
\epsfxsize=6.0cm \epsfysize=6.0cm \centerline{\epsffile{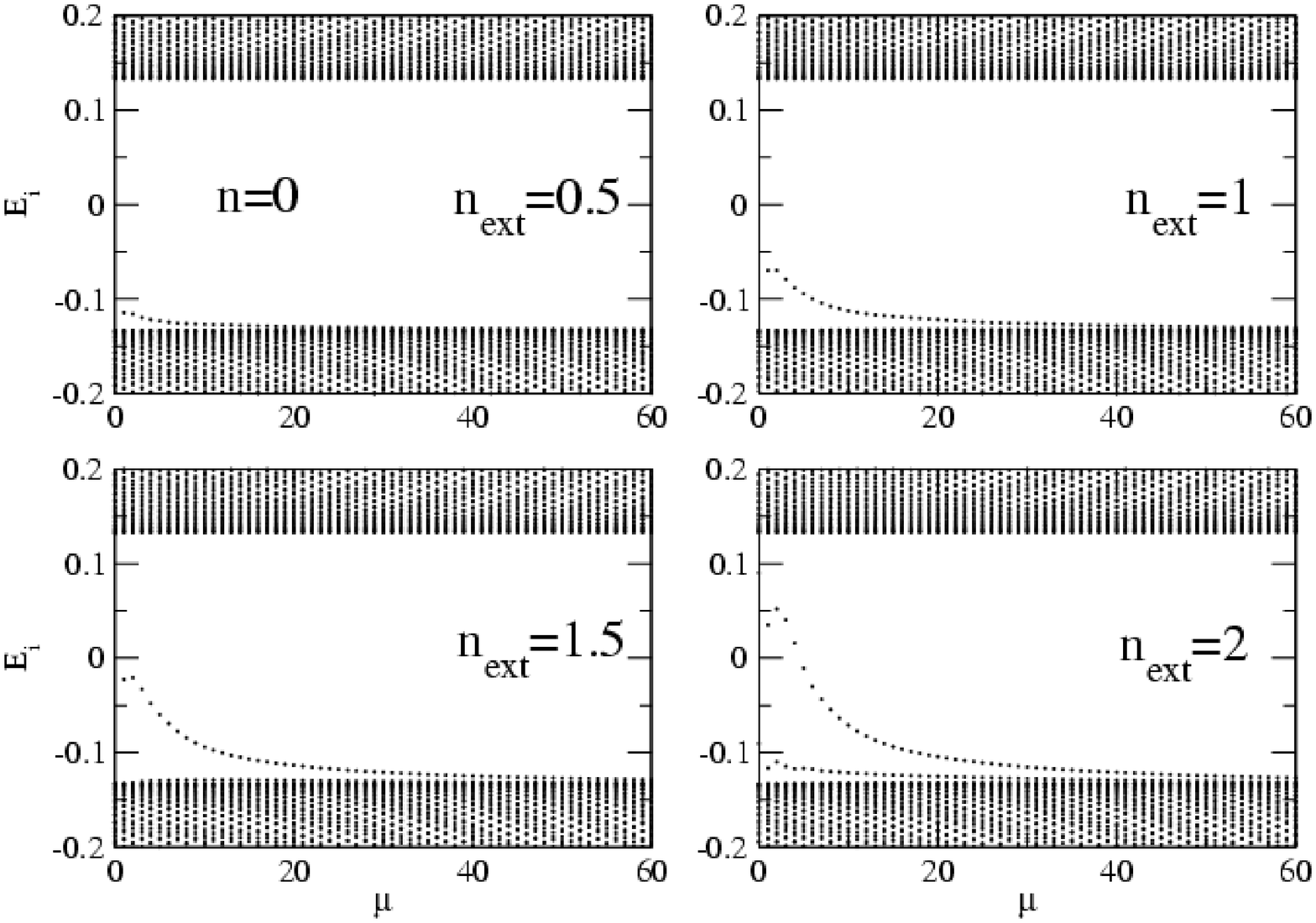}} 
\vspace{1cm} 
\epsfxsize=6.0cm \epsfysize=6.0cm \centerline{\epsffile{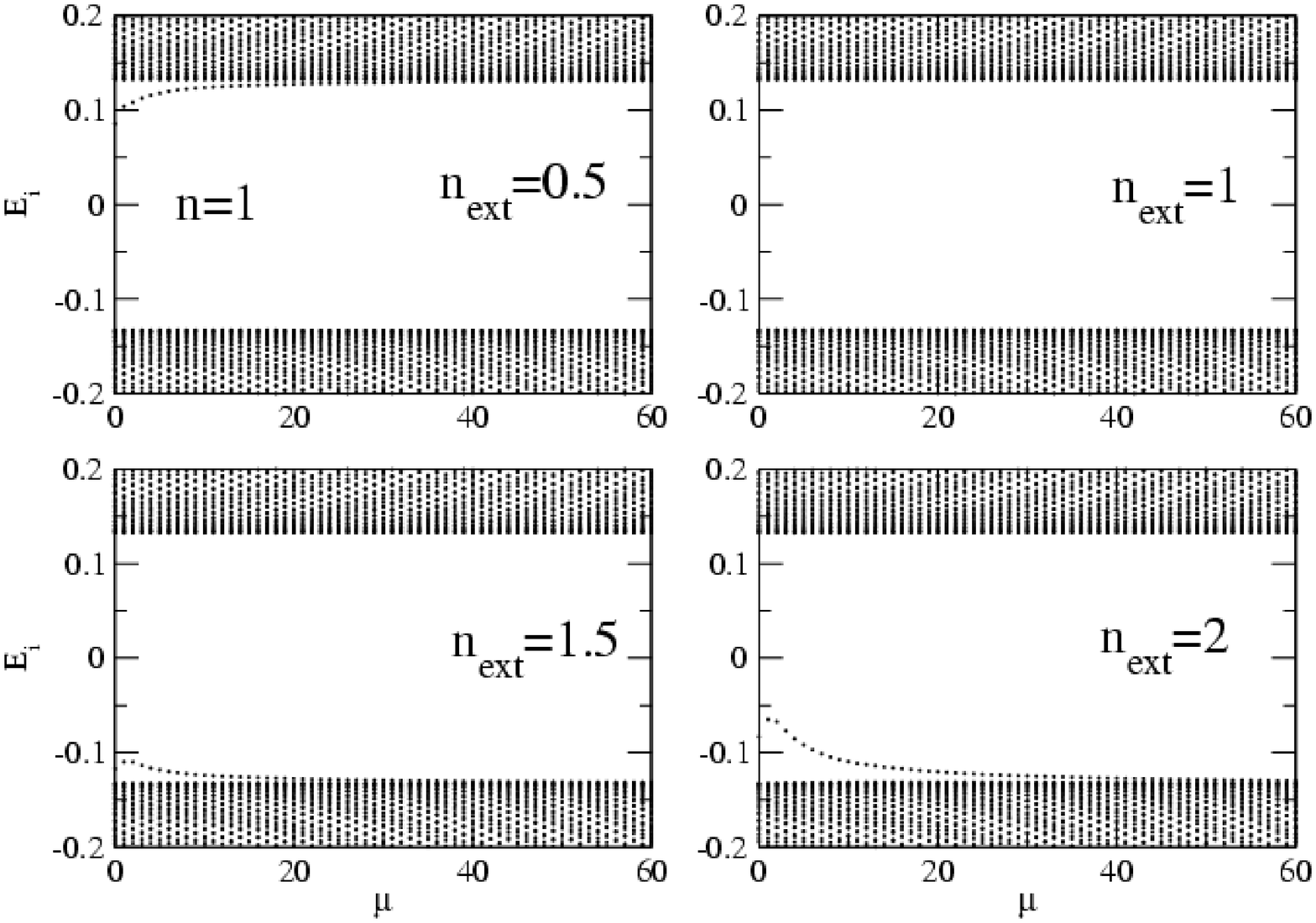}}
\begin{center}
\caption{Energy spectra for (set A) 
(top) different vorticities $n=0, n=1, n=2, n=3$, with $n_{ext}=0$;
(middle) $n=0$ for various external fields $0.5,1.1.5,2$. Note the appearance of negative
energy boundstates as the field increases;
(bottom) $n=1$ for various external fields $0.5,1.1.5,2$. 
Note the appearance of a positive energy
boundstate at a field smaller than the vorticity and negative
energy boundstates as the field increases. Also note the absence of boundstates
for a field $1$.
}
\end{center}
\label{fig4}
\end{figure}

\begin{figure}
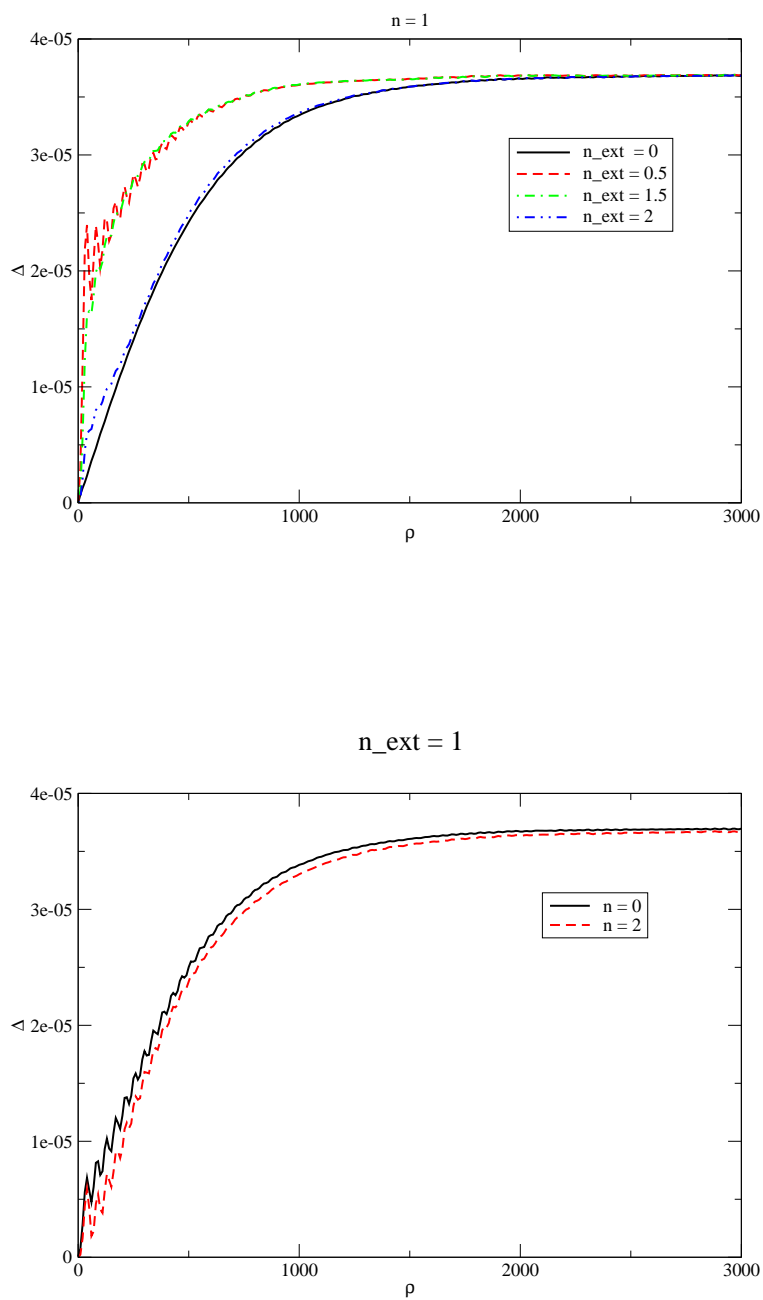

\epsfxsize=10.0cm \epsfysize=10.0cm
\centerline{\epsffile{Fig5a.eps}} \epsfxsize=10.0cm
\epsfysize=10.0cm \centerline{\epsffile{Fig5b.eps}}
\begin{center}
\caption{Gap function for $NbSe_2$ (set B) for (top) $n=1$ and different values of $n_{ext}$ 
and for (bottom) $n_{ext}=1$ and different vorticities $n$.
}
\end{center}
\label{fig5}
\end{figure}

\begin{figure}
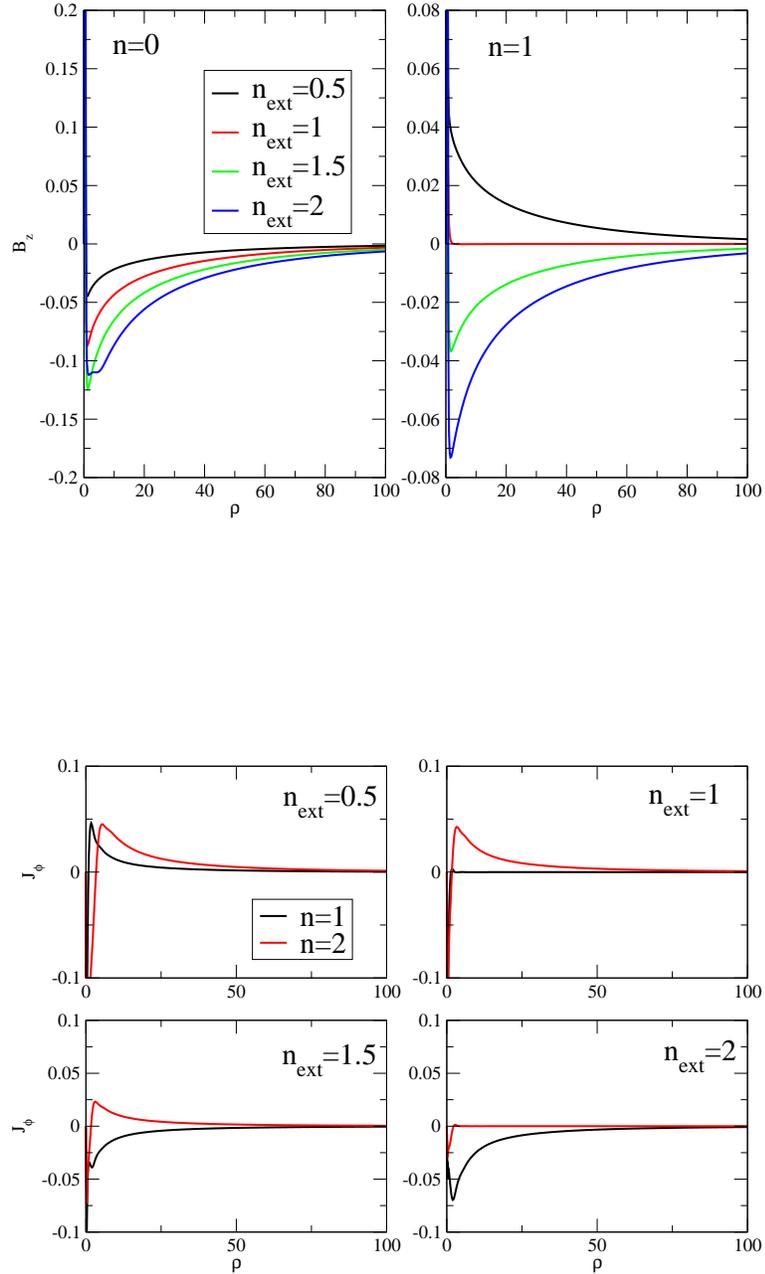

\epsfxsize=10.0cm
\epsfysize=10.0cm
\centerline{\epsffile{Fig6a.eps}} \epsfxsize=10.0cm
\epsfysize=10.0cm \centerline{\epsffile{Fig6b.eps}}
\begin{center}
\caption{a) Magnetic field profiles for the various external field values
for $n=0, n=1$ and for set A. Note the opposing magnetic fields when
$n_{ext} > n $ changes to $n_{ext} < n $. 
b) Induced supercurrents for the various external fields for the
solutions with $n=1,n=2$ for set A. Note the changes of signal close to the
vortex location.
}
\end{center}
\label{fig6}
\end{figure}

In Fig. 4 we show the energy spectra corresponding 
to the cases discussed
in Fig. 3 and compare with the case with no external line ($n_{ext}=0$). 
When the external field is higher 
than the total field and, therefore,
the internal field is negative, there appear boundstates of negative energy in number
$n_{ext}-n$. When the external field is smaller, positive energy boundstates appear
in number $n-n_{ext}$. When the two numbers match there are no bound states, corresponding
to a zero internal field. The boundstates are therefore associated to a non-vanishing
internal field.
Positive energies correspond to a positive internal field and negative energies
to a negative internal field. 
When $n_{ext}$ is an integer the external field can be absorbed in the basis
functions and there is no remaining field affecting the BdG equations (except
for the internal field, which gives a small contribution). This is determined
by $n_a$. If $n_a=0$ there is no field left (external or internal) and this
is equivalent to a system with no vortex. In this case there are no boundstates,
the gap function is uniform and there are no supercurrents.

In Fig. 5 we show the influence of the external field line on the gap
function. The results are presented for the case of $NbSe_2$. When $n_{ext}=0$ the
behavior was previously studied. Since the system is far from the quantum limit
the gap function has a smooth behavior. When the solenoid external field is superimposed,
the shape is altered. As mentioned above the shape of the gap function depends
essentially on the difference $n-n_{ext}$. The coherence length shortens 
which may be possibly interpreted as a lower temperature. Also, the
oscillations increase.

A consistent explanation can be obtained looking at the magnetic field and supercurrent
profiles. In Fig. 6a we compare the magnetic field profiles and in 
Fig. 6b we compare
the supercurrent profiles. In this last case we also consider the case of $n=2$.
As stated above, when the external field is larger than the total field, fixed
by the vorticity of the solution, the magnetic field is negative near the vortex
and tends to zero exponentially from negative values. When the external
field $n_{ext}=n$ the field is zero and when $n_{ext}<n$ the field is positive
as in the case of the vortex line. This is clearly shown in Fig. 6.
The behaviour of the induced supercurrents is more complex. It is also a function
of $n-n_{ext}$. When this difference is zero the current is zero.
When $n-n_{ext}>0$ we have a behaviour similar to the one for the vortex line
with no external field: when $n-n_{ext}=1$ the 
current is positive going through
a maximum and then decreasing, and for larger values of $n-n_{ext}$ the current
has a node. When $n-n_{ext}<0$ the current is always negative.

\begin{figure}
\epsfxsize=8.0cm \epsfysize=6.0cm
\centerline{\epsffile{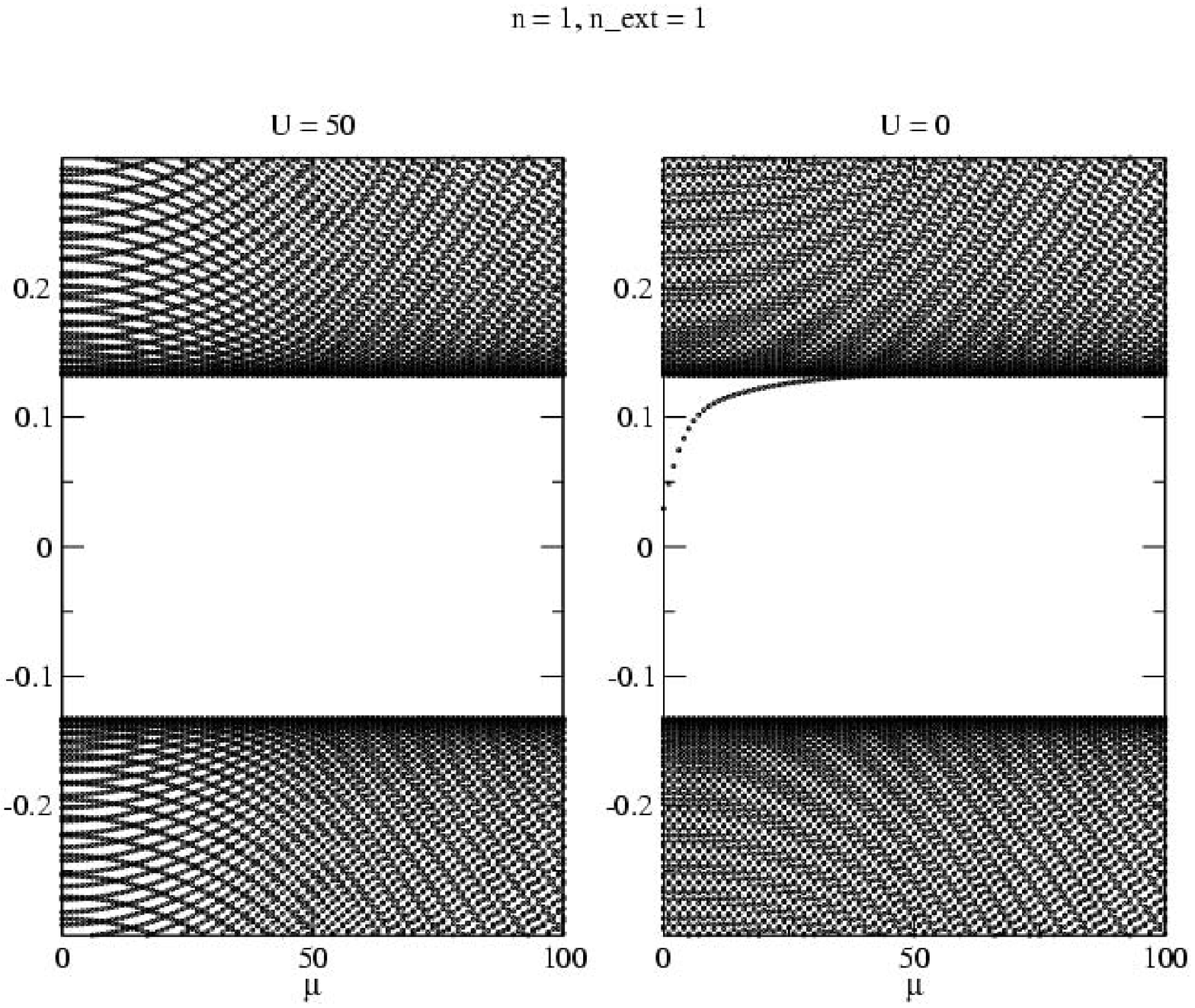}} 
\vspace{1.5cm} \epsfxsize=8.0cm
\epsfysize=6.0cm \centerline{\epsffile{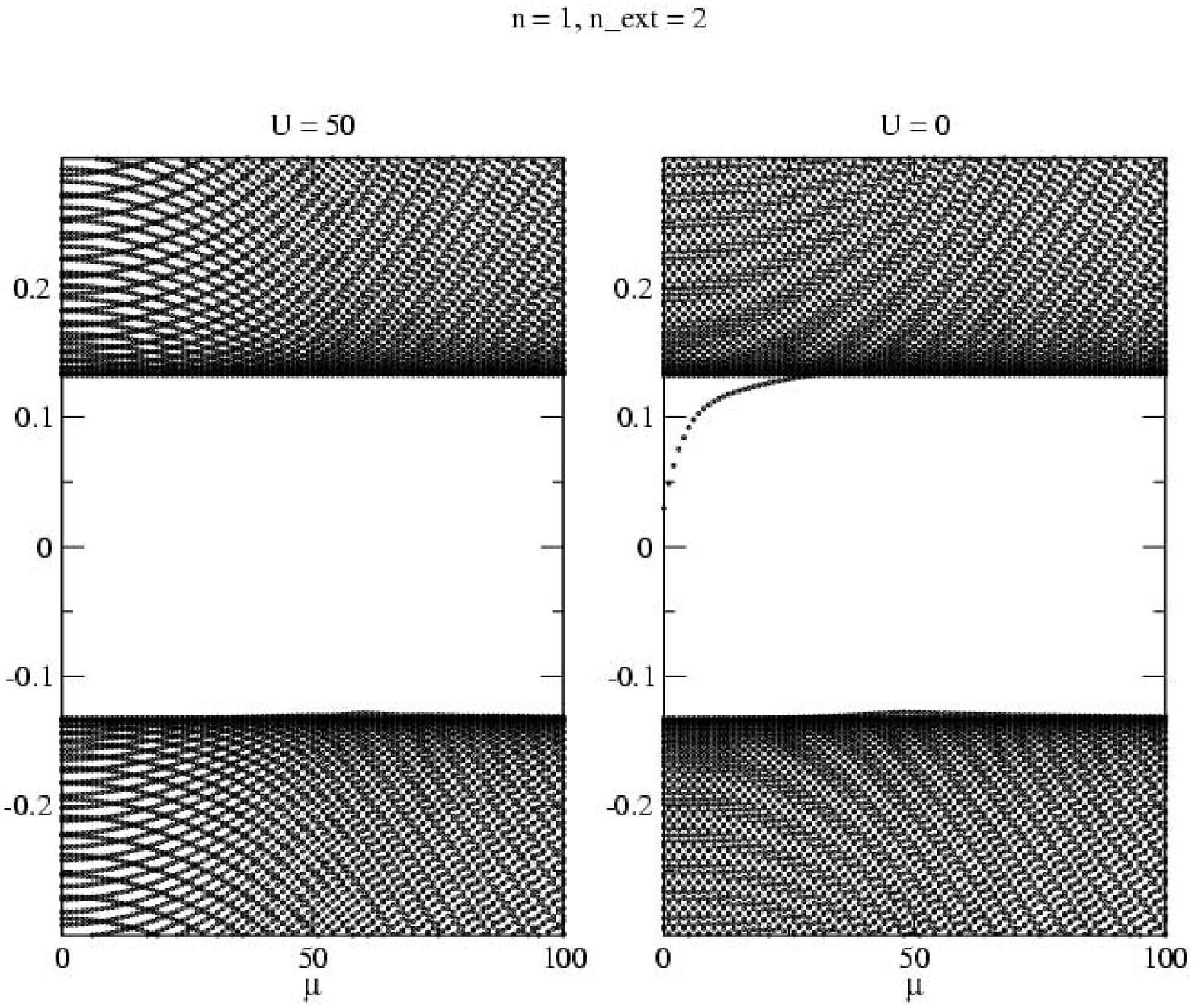}} 
\vspace{1.5cm} \epsfxsize=8.0cm
\epsfysize=6.0cm \centerline{\epsffile{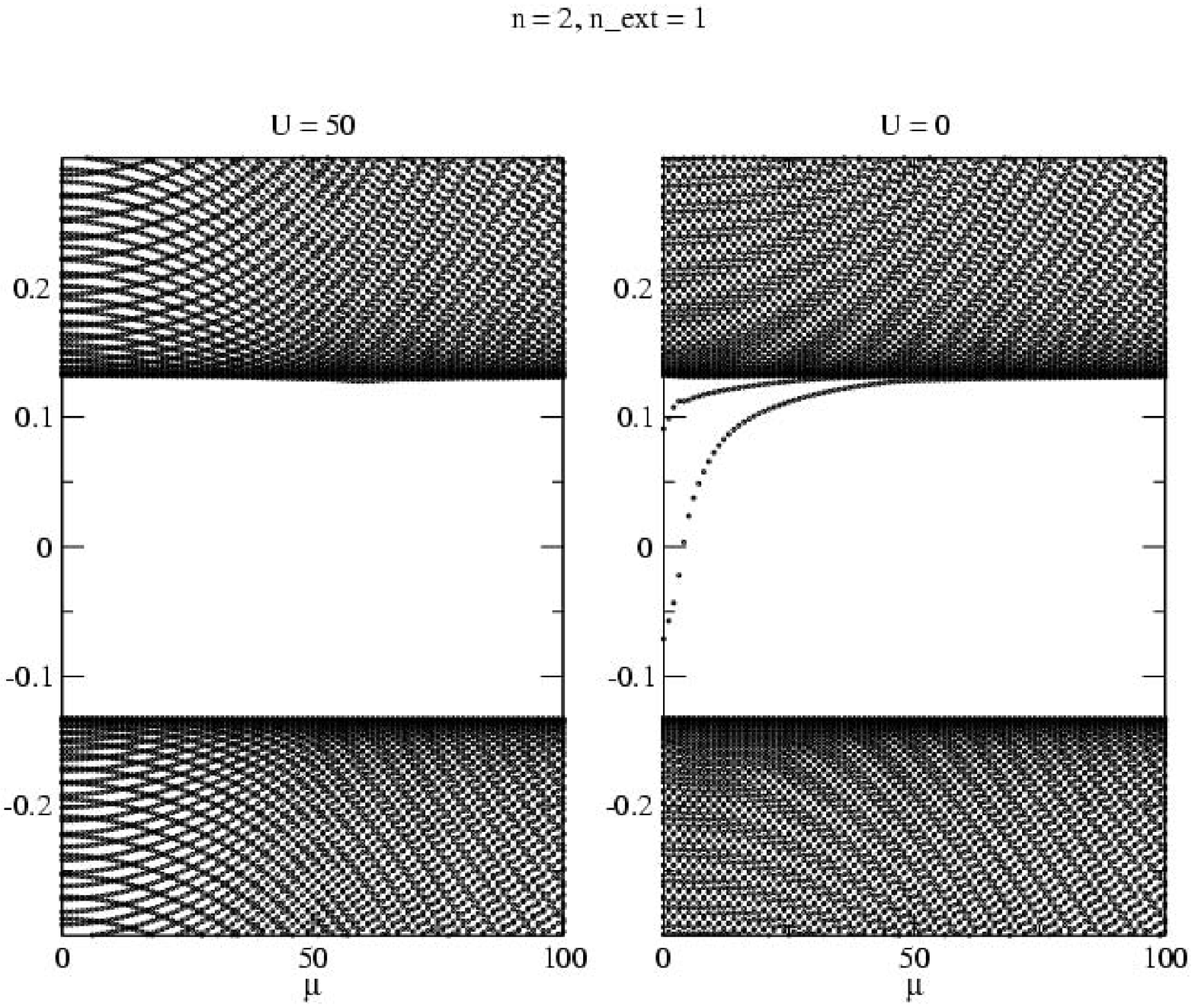}}
\begin{center}
\caption{Energy spectra for two solenoids with and without repulsion 
for different vorticities and external fields for set A.
}
\end{center}
\label{fig7}
\end{figure}

\begin{figure}
\epsfxsize=8.0cm \epsfysize=6.0cm
\centerline{\epsffile{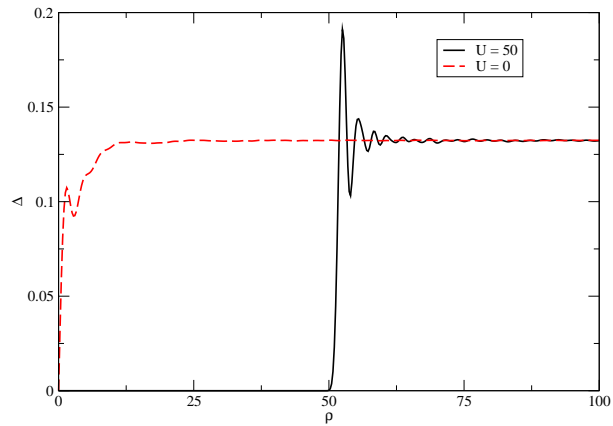}} \vspace{2cm} \epsfxsize=8.0cm
\epsfysize=6.0cm \centerline{\epsffile{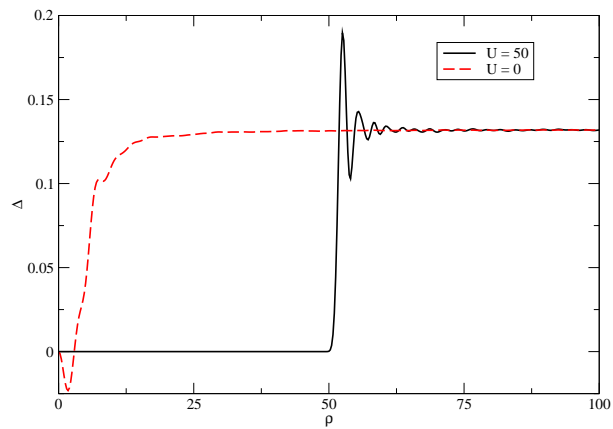}} 
\begin{center}
\caption{Gap function for two solenoids with and without repulsion
for different vorticities and external fields for set A.
}
\end{center}
\label{fig8}
\end{figure}

\subsection{External solenoid of finite width}

Consider now a solenoid of finite width. As mentioned above it could also
be a magnetic whisker inserted in the material. The width of the solenoid
is given by $R_s$. We consider two cases $R_s=1$ and $R_s=50$. The first case
is very similar to the thin solenoid since the coherence length is of the same
order. 
However, the thicker solenoid has a width that is considerably larger
than the coherence length and is of the same order as the penetration length.
The external vector potential is now given by
\be
A_{ext}^{\varphi} = \theta(R_s-\rho) \frac{n_{ext}\Phi_0}{2\pi R_s^2} \rho +
\theta(\rho-R_s) \frac{n_{ext} \Phi_0}{2\pi \rho}
\ee
Also, we consider two cases.  
The first case is achieved considering an external potential
$U(\vec{r})$ that is strongly repulsive in the region occupied by the solenoid.
This effectively restricts the presence of the superconducting pairs in the solenoid.
A value of $U(\mathbf{r})=U \Theta (R_s-\rho)$, with $U=50$ is quite efficient and is
such that the material
inserted in the superconductor is opaque and the electrons can not
penetrate. In the other case we take $U=0$, and therefore the electrons
are allowed to penetrate the finite width region where a magnetic field is inserted
in the superconductor. This allows us to isolate the effect of the finite width region with
magnetic field on the superconductor from the inherent space constraint
due to the physical presence of the solenoid. 

\begin{figure}
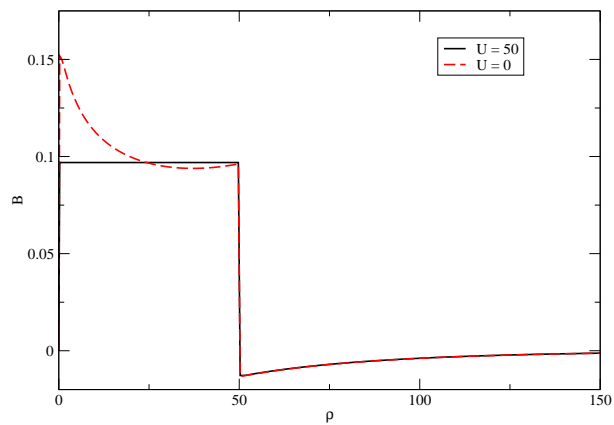
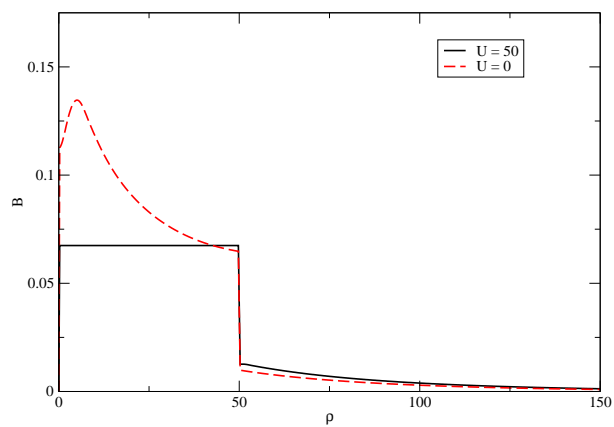

\epsfxsize=8.0cm \epsfysize=6.0cm
\centerline{\epsffile{Fig9a.eps}} \vspace{2cm} \epsfxsize=8.0cm
\epsfysize=6.0cm \centerline{\epsffile{Fig9b.eps}} \vspace{2cm} \epsfxsize=8.0cm
\epsfysize=6.0cm \centerline{\epsffile{Fig9c.eps}}
\begin{center}
\caption{Magnetic field profiles for two solenoids with and without repulsion 
for different vorticities and external fields for set A.
}
\end{center}
\label{fig9}
\end{figure}

\begin{figure}
\epsfxsize=8.0cm
\epsfysize=8.0cm
\centerline{\epsffile{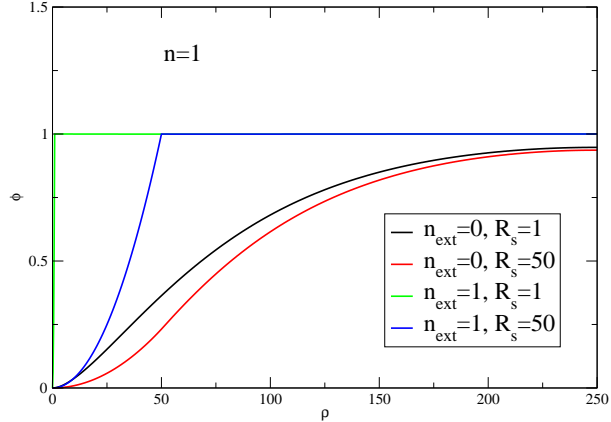}}
\caption{Flux density as a function of distance for $n=1$ and external fields $A_x=0, A_x=1$
for different widths of the opaque solenoid for set A.
}
\label{fig10}
\end{figure}

Consider first the energy spectra. The case of $R_s=1$ is qualitatively similar to the
vortex line and there is a set of boundstates as before characterized by $n_a$.
Even though the vector potential is not singular if the width is very small
there is a large component of the vector potential and the numerical solution
does not distinguish this case from a trully singular potential.
When the thickness is larger
than the coherence length the structure of the boundstates is changed. In the case
of the opaque solenoid the boundstates disappear altogether since the pair density near the
origin vanishes, due to the repulsive potential. If the solenoid is transparent,
the electrons can probe the solenoid core. However, since the magnetic field is spread through
a finite region, the spectrum depends mainly on the vorticity and only weakly
on the external flux. The effect is not strong enough to change the nature of the boundstates.
In this case ther is no singular term and the spectrum is not qualitatively changed.
The number of boundstates is therefore characterized by the vorticity of the gap
function, $n$.
This is shown in Fig. 8 where we compare the energy spectra for the
two cases of $U=50$ and $U=0$. 
The effect of the solenoid on the gap function is shown
in Fig. 9 where the two solenoids are also compared.

The effects of the solenoid are best seen in the magnetic field profiles.
In Fig. 10 we compare the magnetic field for the two solenoids
for different cases. When the vorticity $n<n_{ext}$ the magnetic field
for $\rho>R_S$ is negative, to compensate the field introduced by the
external solenoid. This does not happen when $n>n_{ext}$ since the external
solenoid field does not exceed the field that would correspond to the vorticity
selected. Also, note the difference at small distances between the opaque and
the transparent solenoid. The field at small distances in the opaque solenoid
is determined by $n_{ext}$ which creates a uniform field and determined by the
supercurrents distributed for $\rho>R_s$, which create an opposing uniform
field inside the solenoid.
In the case of the transparent solenoid
the Cooper pairs can penetrate the small distance region and feel the singular
nature of the vortex line implied by the selected form of the gap
function.

In Fig. 11 we show the flux density for the opaque solenoid for 
$n=1$ with and without current
flowing through the solenoid, for the two different widths. 
When $n_{ext}=1$ the
flux saturates in the vicinity of the core for the smaller solenoid. However, in the case
of the larger solenoid the flux just follows the usual classical trend due to the 
constant value of the magnetic field inside the solenoid, assumed of infinite length.  
In the vortex line the field decreases from the core while in the solenoid it is constant.

\section{Conclusions}

In this work we have revisited the problem of a vortex in a superconductor.
There are several reasons that took us to carry out this study. First of
all there has been an increased interest in the interplay of magnetic and
superconducting materials both from the point of view of the influence
of the superconductivity on the magnetic materials but also the influence
of the magnetic materials on the superconductivity. In particular, the vicinity
of magnetic dots near the superconductor may serve as pinning centres and the motion
of vortices in the superconductor may induce the motion of magnetic domain
walls in magnetic materials. These properties may be of importance for the control
of dissipation in the superconductor and the control of magnetic registers
in the magnetic materials, respectively. The studies carried out before
have shown the possibility of observation of giant vortices and therefore we
have considered in this work different vorticities 
and studied the current and gap function profiles.

An interesting problem to be considered is the penetration of a magnetic rod
or solenoid in the bulk of the superconductor. We have considered this
situation and studied the response of the superconductor as a function of the
vorticity around the external field line. 
We have confirmed by a full self-consistent calculation that the winding of the
phase of the gap function determines the properties of the system with the 
notable exception of an infinitely thin solenoid. In this case there is an effective
vorticity, $n_a$, which determines the nature of the spectrum.
We established a detailed connection between the vorticity, the induced internal
currents and the energy spectrum structure.

An extension of this problem
is to consider a superconductor in a finite slab and to insert two
magnetic rods from opposite sides with either the same or opposite
polarities. At the tip of each rod (or solenoid) the magnetic field lines
will either leave or enter the supercondutor. Due to the Meissner effect
these field lines will be confined to flux tubes in a way similar to the
confinement of the chromodynamic field. Far from the solenoid ends
the system looks very much like the problem studied here: a single infinite
solenoid inserted in a superconductor. The behaviour near the solenoid tips
is quite interesting for it provides an explicit realization of the
confinement problem. This problem will be considered elsewhere.

\ack

This work has been partially supported (P.D.S.) through
ESF Science Program INSTANS 2005-2010.


\end{document}